\documentclass[preprint,pre,superscriptaddress,showpacs,showkeys]{revtex4}

\usepackage{graphicx}
\usepackage{amsmath}
\usepackage{amssymb}
\usepackage{amsfonts}

\begin{document}

\title{Two-dimensional flow of foam around an obstacle: force measurements}

\author{Benjamin Dollet}
\affiliation{Laboratoire de Spectrom\'etrie Physique, BP 87, 38402
Saint-Martin-d'H\`eres Cedex, France} \altaffiliation{UMR 5588 CNRS and
Universit\'e Joseph Fourier.}
\author{Florence Elias}
\affiliation{Laboratoire des Milieux D\'esordonn\'es et H\'et\'erog\`enes, case
78, 4 place Jussieu, 75252 Paris Cedex 05, France}
\altaffiliation{F\'ed\'eration de Recherche FR 2438 "Mati\`ere et Syst\`emes
Complexes", Paris, France.}
\author{Catherine Quilliet}
\affiliation{Laboratoire de Spectrom\'etrie Physique, BP 87, 38402
Saint-Martin-d'H\`eres Cedex, France}
\author{Christophe Raufaste}
\affiliation{Laboratoire de Spectrom\'etrie Physique, BP 87, 38402
Saint-Martin-d'H\`eres Cedex, France}
\author{Miguel Aubouy}
\affiliation{SI3M, DRFMC, CEA, 38054 Grenoble Cedex 9, France}
\author{Fran\c cois Graner}
\email{graner@spectro.ujf-grenoble.fr} \affiliation{Laboratoire de
Spectrom\'etrie Physique, BP 87, 38402 Saint-Martin-d'H\`eres Cedex, France}
\altaffiliation{UMR 5588 CNRS and Universit\'e Joseph Fourier}

\date{\today}

\begin{abstract}
A Stokes experiment for foams is proposed. It consists in a two-dimensional
flow of a foam, confined between a water subphase and a top plate, around a
fixed circular obstacle. We present systematic measurements of the drag exerted
by the flowing foam on the obstacle, \emph{versus} various separately
controlled parameters: flow rate, bubble volume, bulk viscosity, obstacle size,
shape and boundary conditions. We separate the drag into two contributions, an
elastic one (yield drag) at vanishing flow rate, and a fluid one (viscous
coefficient) increasing with flow rate. We quantify the influence of each
control parameter on the drag. The results exhibit in particular a power-law
dependence of the drag as a function of the bulk viscosity and the flow rate
with two different exponents. Moreover, we show that the drag decreases with
bubble size, and increases proportionally to the obstacle size. We quantify the
effect of shape through a dimensional drag coefficient, and we show that the
effect of boundary conditions is small.
\end{abstract}

\pacs{82.70.Rr, 83.80.Iz, 47.50.+d, 47.60.+i}

\keywords{foam, Stokes experiment, drag, two-dimensional}

\maketitle

\section{Introduction}

Liquid foams, like colloids, emulsions, polymer or surfactant solutions, are
characterised by a complex mechanical behaviour. These systems, known as soft
complex systems, are multiphasic materials. Their constitutive entities are in
interaction, generating internal structures, which cause diverse rheological
behaviour \cite{Larson1999}. Liquid foams are convenient model experimental
system for studying the interplay between structure and rheology, since their
internal structure can be easily visualised and manipulated.

Liquid foams are made of polyhedral gas bubbles separated by thin liquid
boundaries forming a connected network. The liquid phase occupies a small
fraction of the volume of the foam (a few percent for a dry foam). The
deformations and motions of liquid foams are very diverse: foams are elastic,
plastic or viscous depending on the applied strain and strain rate
\cite{Weaire1999}. This behaviour has been shown in rheological experiments
performed on three-dimensional (3D) foams
\cite{Mason1995,Mason1996,Cohen-Addad1998,Saint-Jalmes1999}; models have been
built to account for this diversity of rheological behaviour
\cite{Reinelt2000,Sollich1997,Durian1995,Cates2004,Hohler2004}. However, the
visualisation of the foam structure is technically difficult in 3D
\cite{Monnereau1999,Prause1995}, although progresses have been made recently
\cite{Lambert2004}. Moreover, the drainage of the liquid phase due to gravity
may occur in 3D, making the fluid fraction and therefore the rheological moduli
of the foam, as well as bubble size (through coarsening) inhomogeneous
\cite{Koehler2001}.

For all these reasons, the mechanics of foams has been studied in two
dimensions, where the direct visualisation of the structure is easier, and no
gravity-driven drainage occurs if the system is horizontal. The system is then
either a true 2D system (unlike bubble raft \cite{Abdelkader1998,Pratt2003}),
like a Langmuir foam \cite{Losche1983,Courty2003}, or a quasi 2D system
constituted by a monolayer of bubbles, either confined between two horizontal
transparent plates (Hele-Shaw cells
\cite{Smith1952,Debregeas2001,Asipauskas2003}: incompressible foams, see below)
or between the surface of the solution and an upper horizontal transparent
plate \cite{Smith1952,Vaz1997} (compressible foams, see below). The deformation
and motion of individual cells have been forced and studied in different flow
geometries: simple shear \cite{Abdelkader1998}, flow in a constriction or
around an obstacle \cite{Asipauskas2003}, Couette flow
\cite{Debregeas2001,Pratt2003}. Some authors have been particularly interested
in the dynamics of bubble rearrangements during the flow: the spatial
distribution of the rearrangements \cite{Abdelkader1998,Debregeas2001}, the
stress relaxation associated with the rearrangements \cite{Pratt2003}, the
deformation profile \cite{Janiaud}, the averaged velocity
\cite{Debregeas2001,Asipauskas2003}. However, no mechanical measurement has
been performed in those last studies.

In this paper, we study the mechanics of a foam flowing in relative
displacement with respect to an obstacle, at a constant velocity. In a
Newtonian liquid at low Reynolds number, the force would vary linearly with the
foam-obstacle relative velocity, the proportionality factor being linked to the
liquid viscosity and the size of the obstacle. This experiment gives
information on the effective viscosity of a flowing foam. Such a Stokes
experiment has first been performed in a 3D coarsening foam by Cox \emph{et
al.} \cite{Cox2000}. Here, we measure the force exerted by the quasi 2D foam on
the obstacle, as a function of the flow velocity, in a 2D geometry. A similar
experiment has been performed recently to investigate the elastic regime of a
2D foam and measure the foam shear modulus \cite{Courty2003}. In the
experiments presented here, the foam flows permanently around the obstacle, and
the stationary regime is investigated. The system used is a monolayer of soap
bubbles confined between the surface of the solution and a horizontal plate.
This allows measuring accurately the forces exerted on the obstacle (section
\ref{ObstacleForce}), and varying easily the foam internal parameters such as
the viscosity of the solution, the bubble size, and the geometry of the
obstacle.

The article is organised as follows. The experimental materials and methods are
presented in section II, and the results are shown in section III. These
results are discussed in section IV, and conclusions are exposed in section V.

\section{Materials and methods}

\subsection{Foam production}

\begin{figure}
\begin{center}
\includegraphics[width=8cm]{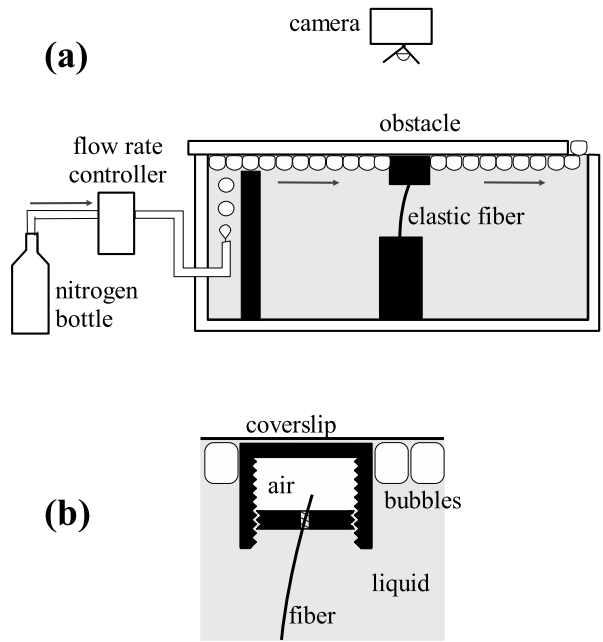}
\caption{\label{Setup+flotteur} (a) Experimental setup. The arrows indicate the
flow of gas and foam. (b) Detailed sketch of the obstacle.}\end{center}
\end{figure}

\begin{figure}
\begin{center}
\includegraphics[width=8cm]{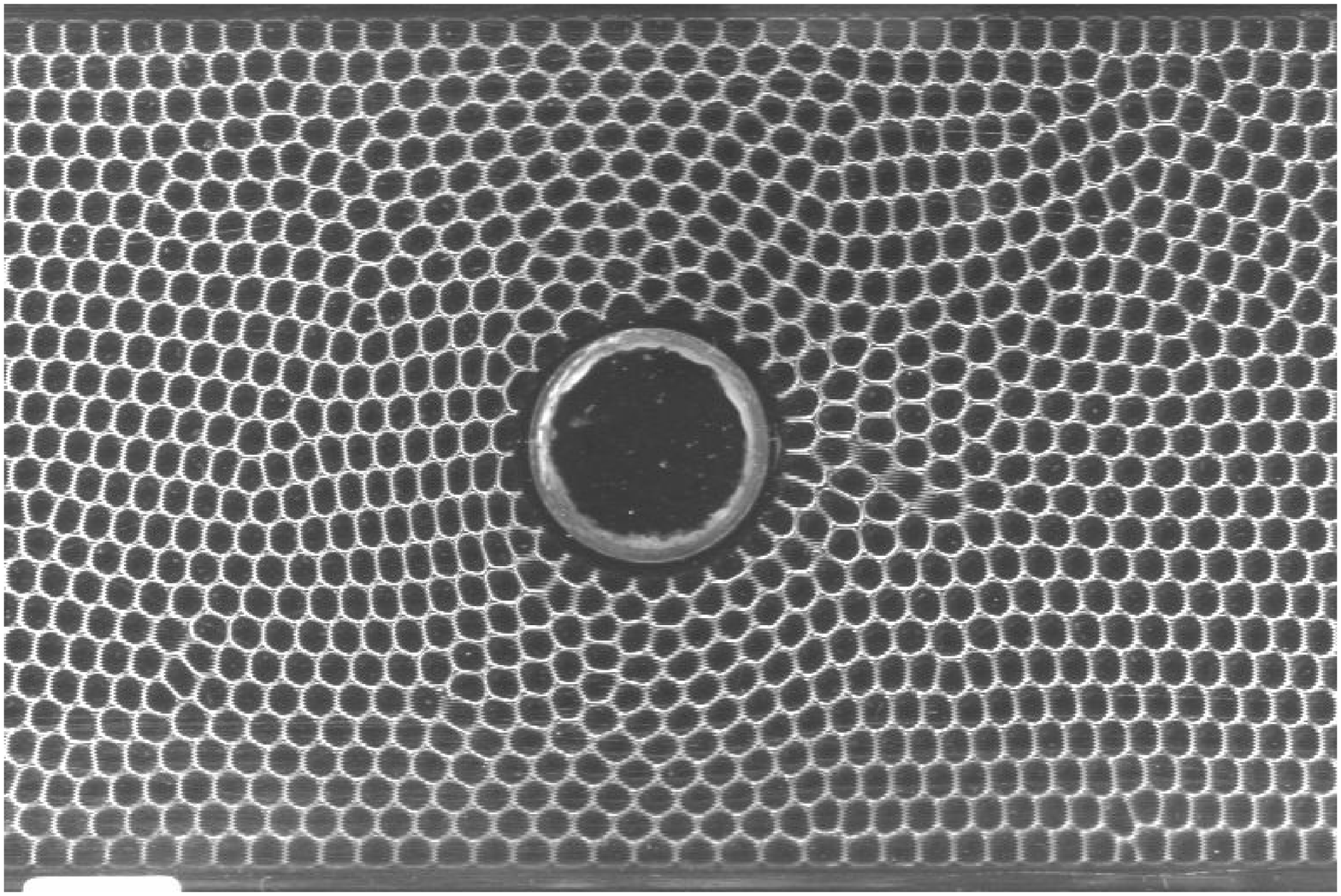}
\caption{\label{Image} Photo of foam flowing from left to right around a
circular obstacle of diameter 30 mm. The bubble size is 16.0 mm$^2$ (note the
monodispersity of the foam), and the flow rate is 174 ml min$^{-1}$. The walls
of the channel (width 10 cm) are visible at the top and bottom of the picture.
The stretching and shearing of bubbles due to the presence of the obstacle is
clearly visible around the obstacle. The surface of the observed field is
$15.4\times 10.2$ cm$^2$, and 1 pixel side equals 0.20 mm. Films are available
at \texttt{http://www-lsp/link/mousses-films.htm} at low (17 ml/min), moderate
(112 ml/min) and high (515 ml/min) flow rates for this obstacle and bubble
area.}\end{center}
\end{figure}

The experimental setup is presented on Fig. \ref{Setup+flotteur}(a). The
experiments are performed in a glass channel of 110 cm length, 10 cm width and
10 cm depth. The soap solution is a solution of commercial dish-washing fluid
(1\% in volume) in purified water, with added glycerol when the viscosity needs
to be varied (section \ref{SectionViscosity}). The surface tension of the
solution is $\gamma=31\pm 3$ mN m$^{-1}$. At the beginning of each experiment,
the channel is filled with the solution, with a gap of thickness $3.50\pm 0.05$
mm between the liquid surface and the coverslip. The foam is produced by
blowing bubbles of nitrogen in the solution, at one end of the channel, in a
chamber bounded by a barrier which allows a single monolayer of bubbles to
form. The continuous gas flow makes the foam flow along the channel, between
the surface of the solution and the coverslip, until it reaches the open end of
the channel, where bubbles pop in contact with the atmosphere. Leaks are
carefully avoided, so that the total amount of liquid in the channel is
constant during an experiment, and for each experiment. A typical image of the
flowing foam observed from above is displayed in Fig. \ref{Image}.

\subsection{\label{ObstacleForce}Obstacle and force measurements}

The obstacle stands in the middle of the channel. It is a buoyant mobile
plastic cylinder connected to a fixed base by a soft glass fiber. The bottom
extremity of the fiber is rigidly fixed. Its top extremity simply passes
through a hole drilled in the bottom of the cylinder (Fig
\ref{Setup+flotteur}(b)). Therefore, the fiber can slide inside the
horizontally moving cylinder, without applying any undesirable vertical force.
Moreover, the fiber is lubricated by the liquid, which avoids solid friction
against the cylinder.

The horizontal force $F$ exerted by the foam on the obstacle tends to pull it
streamwise; it is balanced by the horizontal drawback force $F_d$ from the
elastic fiber, which deflection is designed by $X$. The calculation of this
force is classical in the theory of elasticity \cite{Landau1986}; since the
deflection of the fiber is too large to use linear Hooke's law, we use the
following one:
\begin{equation}\label{Force}
  F_d = -\frac{\pi ED^4}{64L^2} \left[ 3\frac{X}{L} - \frac{81}{35} \left( \frac{X}{L} \right)^3
  + \frac{29646}{13475} \left( \frac{X}{L} \right)^5 + \mathcal{O}\left( \frac{X}{L} \right)^7 \right] ,
\end{equation}
where $D=240$ $\mu$m is the fiber diameter, $L$ its vertical length and $E$ its
shear modulus. This expansion gives a precision of 0.3\% over the force. The
derivation of formula (\ref{Force}) is detailed in the appendix. The fiber has
been calibrated by measuring its deflection under its own weight, giving the
value of the parameter : $ED^4 = (2.21\pm 0.02)\times 10^{-4}$ Pa m$^4$. This
value is compatible with typical values of the Young modulus of glass:
$6$--$7\times 10^{10}$ Pa. We use two different fibers of vertical lengths:
$L=34.8\pm 0.1$ mm and $L=42.4\pm 0.1$ mm, depending on the magnitude of the
force to measure. We have checked that for given experimental conditions, the
same force measured with both fibers yields the same result (data not shown).
The displacement is measured by tracking the position of the obstacle with a
CCD camera placed above the channel: the actual position of the obstacle is
given by the coordinates of its center, obtained by image analysis. The
position of the center of the obstacle is known with a precision of 0.02 mm,
much lower than the typical displacement (5 mm to 1 cm). When the obstacle has
reached a stationary position under flow, the drawback force exactly
compensates the force exerted by the foam, which is then directly deduced from
the measured displacement.

The obstacle is in contact with the coverslip. This is necessary for the foam
to flow around the obstacle and not above it, but this may induce friction.
Nevertheless, in the setup presented here, the obstacle is in contact with a
single plate; this reduces the friction in comparison with an experiment
performed in a Hele-Shaw cell, where the foam is confined between two plates.
Furthermore, the obstacle has an enclosed cavity closed by a watertight screw
(Fig. \ref{Setup+flotteur}(b)), which enables to tune its buoyancy such that
the contact force with the top plate is minimal. In the presence of the foam,
the obstacle is in contact with the top plate through a capillary bridge,
avoiding solid friction. We check for each experiment that the obstacle is not
stuck: its position fluctuates under the slight flow heterogeneities, and
results presented below average the position of the obstacle over 50 successive
images with an interval of two seconds. Viscous friction between the obstacle
and the coverslip cannot be eliminated, but it only influences transients,
which are not considered in this paper: each measurement is performed in a
stationary regime. Reversibility and reproducibility tests give an upper bound
for the force measurement errors: 0.2 mN, to be compared to the typical forces,
of the order of 5 mN.

As shown by Fig. \ref{Setup+flotteur}(b), a part of the obstacle is immersed in
the subphase, which may be drawn by the flowing foam. This flowing subphase
exerts an additional force on the obstacle, which is negligible as shown by the
following evaluation. The total height of the obstacle is 23 mm, so the
immersed height is $h \simeq 19.5$ mm because the foam thickness remains close
to the initial thickness of 3.5 mm between the solution and the coverslip.
Therefore, a generous upper bound of the drag exerted by the subphase would be
obtained by assuming that the subphase flows at the same velocity $V$ that the
foam. The diameter of the obstacle being $2R=30$ mm and the width of the
channel $2H=10$ cm, the drag exerted by the flowing subphase of dynamic
viscosity $\eta$ would equal \cite{Faxen1946} :
\begin{equation}\label{Faxen}
F_{\mathrm{subphase}} \simeq \frac{4\pi\eta hV}{\ln H/R - 0.91} .
\end{equation}
By taking the highest foam velocity reached in the experiments $V=3$ cm
s$^{-1}$ and the highest dynamic viscosity used: $\eta=9\times 10^{-3}$ Pa s,
the upper bound of the force would be then evaluated to $F_{\mathrm{subphase}}
= 0.2$ mN, which is comparable to the other sources of error, and much lower
than the typical forces exerted by the foam on the obstacle.

\subsection{\label{ControlParameters}Control parameters}

A first control parameter is the nitrogen flow rate $Q$, which is adjusted
using an electronic controller (Brooks Instrument B.V.) driven by a home-made
software. The range of available flow rate runs on more than three decades,
from 1 to 2,000 ml min$^{-1}$, with a precision of 0.1 ml min$^{-1}$. Another
control parameter is the bubble volume. It is indirectly determined by
measuring the surface density of bubbles against the coverslip thanks to image
analysis, using NIH Image software. Since the mean foam thickness is fixed by
the total amount of liquid in the channel, which is carefully kept constant,
there is a unique relation between the bubble volume and the mean surface
density. Instead of this surface density, we will refer throughout this paper
to its inverse, that we shall call the mean bubble area. This parameter
slightly differs from the bubble area one can measure directly on an image,
because it includes the water contained in the films and Plateau borders
surrounding bubbles. In our setup, contrary to Hele-Shaw cells, the depth of
the bubbles is free to adjust to pressure variations; this entails an effective
compressibility of the flow, and local variations of bubble area near the
obstacle, as we shall see later (section \ref{DiscussionArea}). The surface
density is measured at the left extremity of the observed field, where the
influence of the obstacle is not significant (Fig. \ref{AreaField}).

For a given injector, the bubble volume increases with the gas flow rate. To
control these two parameters separately, we blow the gas through one to six
tubes (or needles) of same diameter simultaneously, keeping constant the flow
rate per tube, hence the bubble volume. Furthermore, the diameter of these
injectors can be varied, which changes the flow rate per tube for the same
bubble area; hence, for a given bubble volume, typically ten different values
of flow rate are available (from 5 to 13 in the following data), with greatest
flow rate at least 20 times greater than the lowest one. In this paper, we
always produce monodisperse foams: the bubble area disorder, measured as the
ratio of the standard deviation with the mean value of the bubble area
distribution, is lower than 5\%. Six different bubble areas were used: 12.1,
16.0, 20.0, 25.7, 31.7 and 39.3 mm$^2$, chosen with a relative precision of
3\%. The study of smaller bubbles would be problematic, since a transition from
bubble monolayer to multilayer occurs at low bubble width/height ratio
\cite{Cox2002}. At the other extremity, we cannot make a monodisperse foam with
larger bubbles.

Another tunable parameter is the viscosity of the solution, that we will call
bulk viscosity throughout the text. We control it by adding glycerol to the
initial soap solution. We have used five different solutions, with 0, 20, 30,
40 and 50\% glycerol in mass. The respective kinematic viscosities $\nu$,
measured with a capillary viscometer (Schott-Ger\"ate) at room temperature, are
equal to 1.06, 1.6, 2.3, 3.8 and 9.3 mm$^2$ s$^{-1}$. The variation of
viscosity due to the variation of room temperature is lower than 4\%.

\begin{figure}
\begin{center}
\includegraphics[width=7cm]{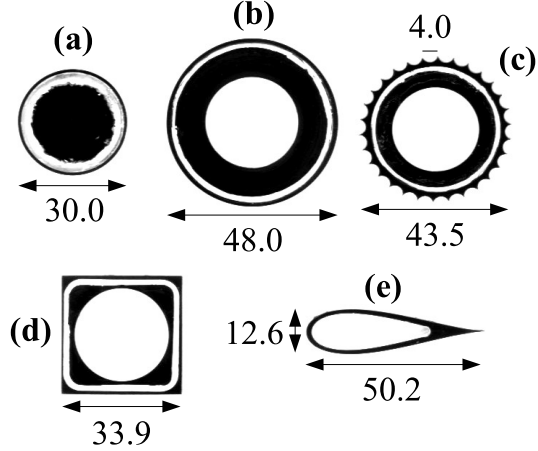}
\caption{\label{Obstacles} Top views of the five obstacles, with dimensions in
millimeters.}\end{center}
\end{figure}

Different obstacles have been used (Fig. \ref{Obstacles}). To change the
obstacle, additional profiles are fixed on the previously described cylinder;
for each obstacle, the apparent density is tuned to avoid solid friction
(section \ref{ObstacleForce}). Two different cylinders of diameter 30.0 (Fig.
\ref{Obstacles}(a)) and 48.0 mm (Fig. \ref{Obstacles}(b)) are used to study the
influence of size. Boundary conditions on the obstacle are investigated using a
cogwheel of diameter 43.5 mm, with circular cogs of diameter 4.0 mm (Fig.
\ref{Obstacles}(c)): whereas flowing foam slips along any smooth obstacle, the
cogs trap the first layer of bubbles surrounding the cogwheel. A square
obstacle, of side 33.9 mm (Fig. \ref{Obstacles}(d)), is used to study
orientation effects. Furthermore, we made an airfoil profile (Fig.
\ref{Obstacles}(e)) to study possible streamlining. It is a standard NACA 0025
profile, which means that it is not cambered and that its maximal thickness
(12.6 mm) equals 25\% of its total length (50.2 mm). This profile was home-made
using a numerical milling machine (Deckel-Maho); its mathematical expression,
parameterized by the angle $t$ running from $-\pi$ to $\pi$, writes: $x(t) =
25.1\cos t$, $y(t) = 4.83(1+\cos t)\sin t$, where the lengths are expressed in
millimeters.

\section{Results}

\subsection{\label{SectionViscosity}Influence of bulk viscosity}

We study the variation of the drag \emph{versus} the flow rate and the bulk
viscosity, for the five different viscosities indicated in section
\ref{ControlParameters}. All these measurements are performed at a fixed bubble
area of 20 mm$^2$, and we use a circular obstacle of diameter 30 mm (Fig.
\ref{Obstacles}).

\begin{figure}
\begin{center}
\includegraphics[width=7cm]{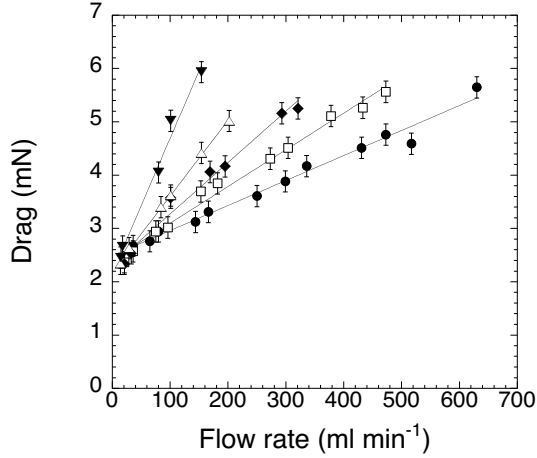}
\caption{\label{Drag (viscosity)} Drag \emph{versus} flow rate, for bulk
viscosity equal to 1.06 ($\bullet$), 1.6 ($\square$), 2.3 ($\blacklozenge$),
3.8 ($\vartriangle$) and 9.3 mm$^2$ s$^{-1}$ ($\blacktriangledown$). The
straight lines are linear fits of the data. The bubble area is 20 mm$^2$ and
the obstacle is a circle of diameter 30 mm. }\end{center}
\end{figure}

\begin{figure}
\begin{center}
\includegraphics[width=7cm]{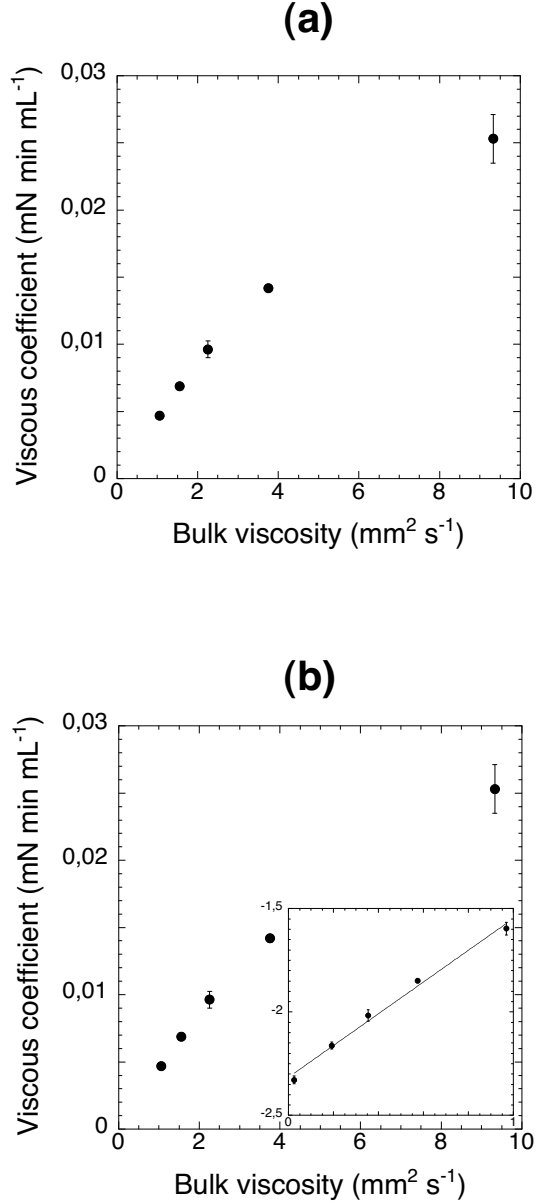}
\caption{\label{Viscosity} Results from fits to Fig. \ref{Drag (viscosity)}.
(a) Yield drag \emph{versus} the bulk viscosity (semi-logarithmic scale), and
(b) viscous coefficient \emph{versus} the bulk viscosity (linear scale).
Insert: log-log plot. All error bars indicate the incertitude on the fit
parameter arising from statistical dispersion of the data. The straight line is
the linear fit: its slope is $0.77 \pm 0.05$.}\end{center}
\end{figure}

We observe two general features (Fig. \ref{Drag (viscosity)}), independent of
the value of the bulk viscosity: the drag does not tend to zero at vanishing
flow rate, and it increases with flow rate. The first observation is a
signature of the solid-like properties of the foam. The second feature is
related to the fluid-like properties of the foam. The data are well fitted by a
linear law (Fig. \ref{Viscosity}):
\begin{equation}\label{Linear fit}
  F=F_0+mQ.
\end{equation}
We call $F_0$ the yield drag, as a reference to the yield properties of the
foam, and the slope $m$ the viscous coefficient, since we can dimensionally
deduce from $m$ an effective 3D viscosity $\mu$ for the foam: $\mu\approx
mS/R$, where $S$ is the cross-section of the foam, and $R$ is the typical size
of the obstacle. Yield drag \emph{versus} bulk viscosity is plot on Fig.
\ref{Viscosity}(a), and viscous coefficient \emph{versus} bulk viscosity on
Fig. \ref{Viscosity}(b).

Fig. \ref{Viscosity}(a) shows that the yield drag is essentially independant of
the bulk viscosity. This was expected, because yield drag is only related to
the yield properties of the foam, which depend on surface tension and bubble
size \cite{Princen1983}. The slight decrease with the bulk viscosity is likely
due to a slight decrease of surface tension with the concentration of the
glycerol, which has been observed for pure water-glycerol solutions. For
example, an equal mixture of water and glycerol lowers the surface tension by
about 10\% \cite{Handbook2003}.

Fig. \ref{Viscosity}(b) shows that the viscous coefficient increases with the
bulk viscosity. The data can be fitted by a power law (insert of Fig.
\ref{Viscosity}(b)), that yields the following dependency of viscous
coefficient on bulk viscosity: $m \propto \nu^{0.77\pm 0.05}$, the error bar
being obtained by the statistical dispersion of the data in the insert of Fig.
\ref{Viscosity}(b).

\subsection{\label{SectionArea}Influence of bubble area}

We now turn to the study of drag versus flow rate and bubble area. All the
measurements are done without adding glycerol in the solution, at a constant
viscosity of 1.06 mm$^2$/s. The obstacle is a cylinder of radius 30 mm. We
study the six bubble areas indicated in section \ref{ControlParameters}, from
12.1 mm$^2$ to 39.3 mm$^2$.

\begin{figure}
\begin{center}
\includegraphics[width=7cm]{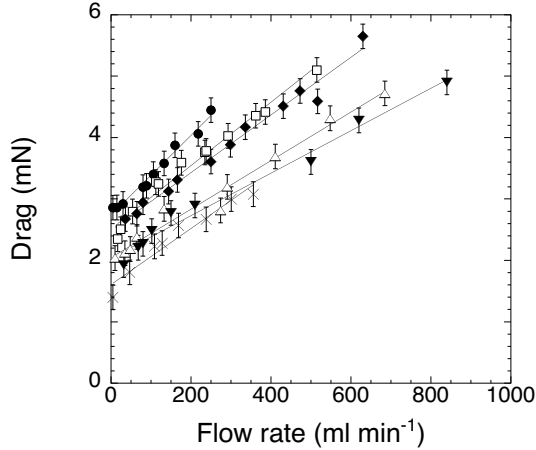}
\caption{\label{Drag (area)} Drag \emph{versus} flow rate, for bubble area
equal to 12.1 ($\bullet$), 16.0 ($\square$), 20.0 ($\blacklozenge$), 25.7
($\vartriangle$), 31.7 ($\blacktriangledown$) and 39.3 mm$^2$ ($\times$). The
straight lines are linear fits of the data. The bulk viscosity is 1.06 mm$^2$
s$^{-1}$ and the obstacle is a circle of diameter 30 mm.}\end{center}
\end{figure}

\begin{figure}
\begin{center}
\includegraphics[width=7cm]{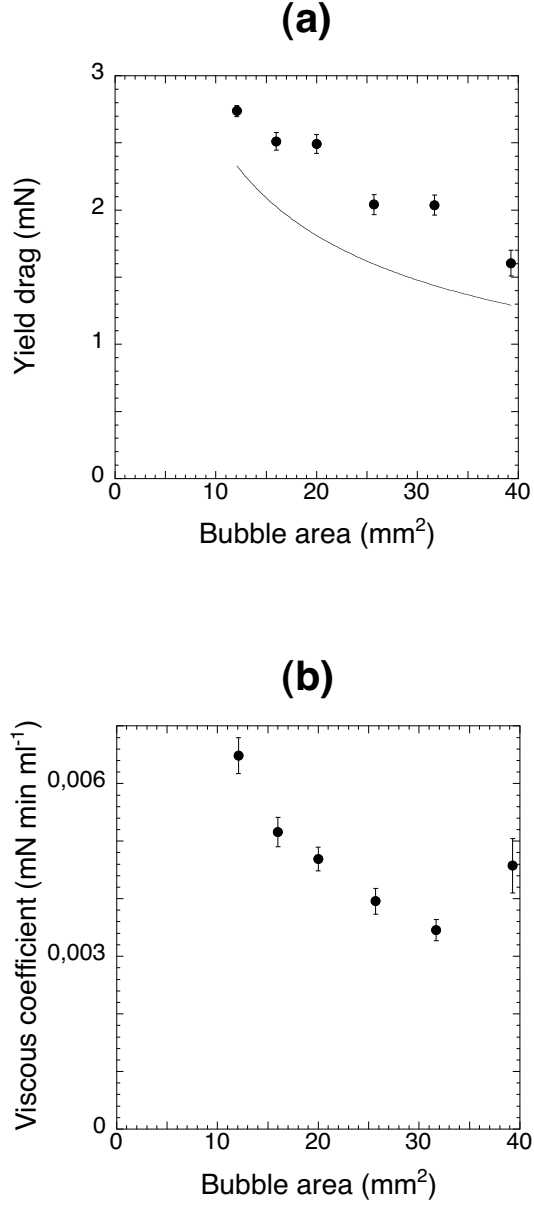}
\caption{\label{Area} Results from fits to Fig. \ref{Drag (area)}. (a) Yield
drag \emph{versus} bubble area. The curve is an evaluation of the elastic
contribution to the drag, see section \ref{DiscussionArea}; (b) viscous
coefficient \emph{versus} bubble area.}\end{center}
\end{figure}

We find again the signature of the viscoelastic properties of the foam (Fig.
\ref{Drag (area)}), with a non-zero yield drag and an increase of drag
\emph{versus} flow rate. We perform again a linear fit (\ref{Linear fit}),
despite a slight non-affine variation for 39.3 mm$^2$, and get the yield drag
and the viscous coefficient, plotted \emph{versus} bubble area in Fig.
\ref{Area}.

Fig. \ref{Area}(a) evidences that the yield drag is a decreasing function of
the bubble area. This is coherent with the fact that both quantities used to
describe the solid properties of the foam, its shear modulus and yield stress,
are also decreasing functions of the bubble size
\cite{Princen1985,Mason1995,Mason1996}. Fig. \ref{Area}(b) shows that the
viscous coefficient is also a decreasing function of bubble area, except for
the last point. The data will be discussed in more detail in section
\ref{DiscussionArea}.

\subsection{\label{SectionObstacle}Influence of obstacle geometry}

We now study a third control parameter, the obstacle geometry, using the five
obstacles described in section \ref{ControlParameters}. As in the previous
section, the solution of viscosity of 1.06 mm$^2$ s$^{-1}$ is used. A bubble
area of 16.0 mm$^2$ was chosen to ensure an optimal trapping of the bubbles in
the cogs of the cogwheel. We focus successively on the influence of
orientation, size, shape and boundary conditions of the obstacles.

\subsubsection{Orientation}

Because of their symmetry, the cylinders and the cogwheel do not display any
orientation effect. We thus focus on the influence of the orientation relative
to the flow direction of the square on the drag measurements.

\begin{figure}
\begin{center}
\includegraphics[width=7cm]{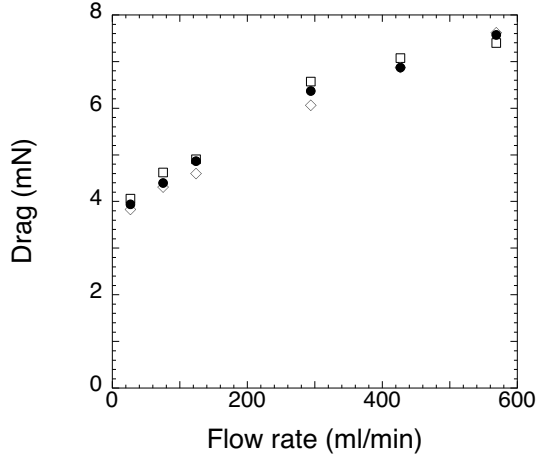}
\caption{\label{Drag_on_square(orient)} Drag on the square obstacle
\emph{versus} flow rate, for orientation equal to 0$^\circ$ ($\bullet$),
22.5$^\circ$ ($\square$) and 45$^\circ$ ($\lozenge$). The bulk viscosity is
1.06 mm$^2$ s$^{-1}$ and the bubble size is 16 mm$^2$.}\end{center}
\end{figure}

We have checked that for the square obstacle, any given orientation is stable.
More precisely, orientation drifts under 90 minutes are always less than
5$^\circ$ (data not shown), although it is a much longer duration than what is
required for the measurements. We have studied the variation of drag
\emph{versus} flow rate for three orientations of the square between a side and
the flow direction: 0$^\circ$, 22.5$^\circ$ and 45$^\circ$. Fig.
\ref{Drag_on_square(orient)} shows that the drag does not depend significantly
on the orientation: henceforth, drag measurements on the obstacle will be
averaged over these three orientations.

Contrary to the circle, the airfoil only possesses two stable orientations,
when its plane of symmetry is parallel to the flow direction. The more stable
configuration is obtained when foam flows from the rounded leading edge to the
sharp trailing edge, which is the usual configuration in aerodynamics.

\subsubsection{Size, shape and boundary conditions}

\begin{figure}
\begin{center}
\includegraphics[width=8cm]{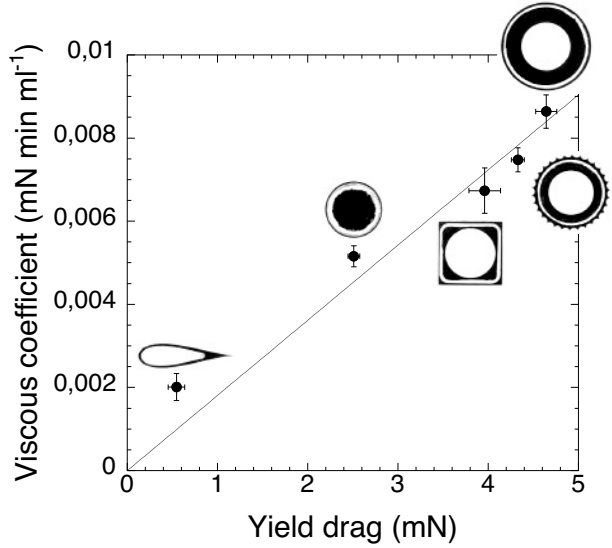}
\caption{\label{Viscous_coeff(yield_drag)} Viscous coefficient \emph{versus}
yield drag for the five obstacles, whose photos are sketched near the
corresponding data. The straight line is the linear fit passing through zero of
the data.}\end{center}
\end{figure}

\begin{figure}
\begin{center}
\includegraphics[width=7cm]{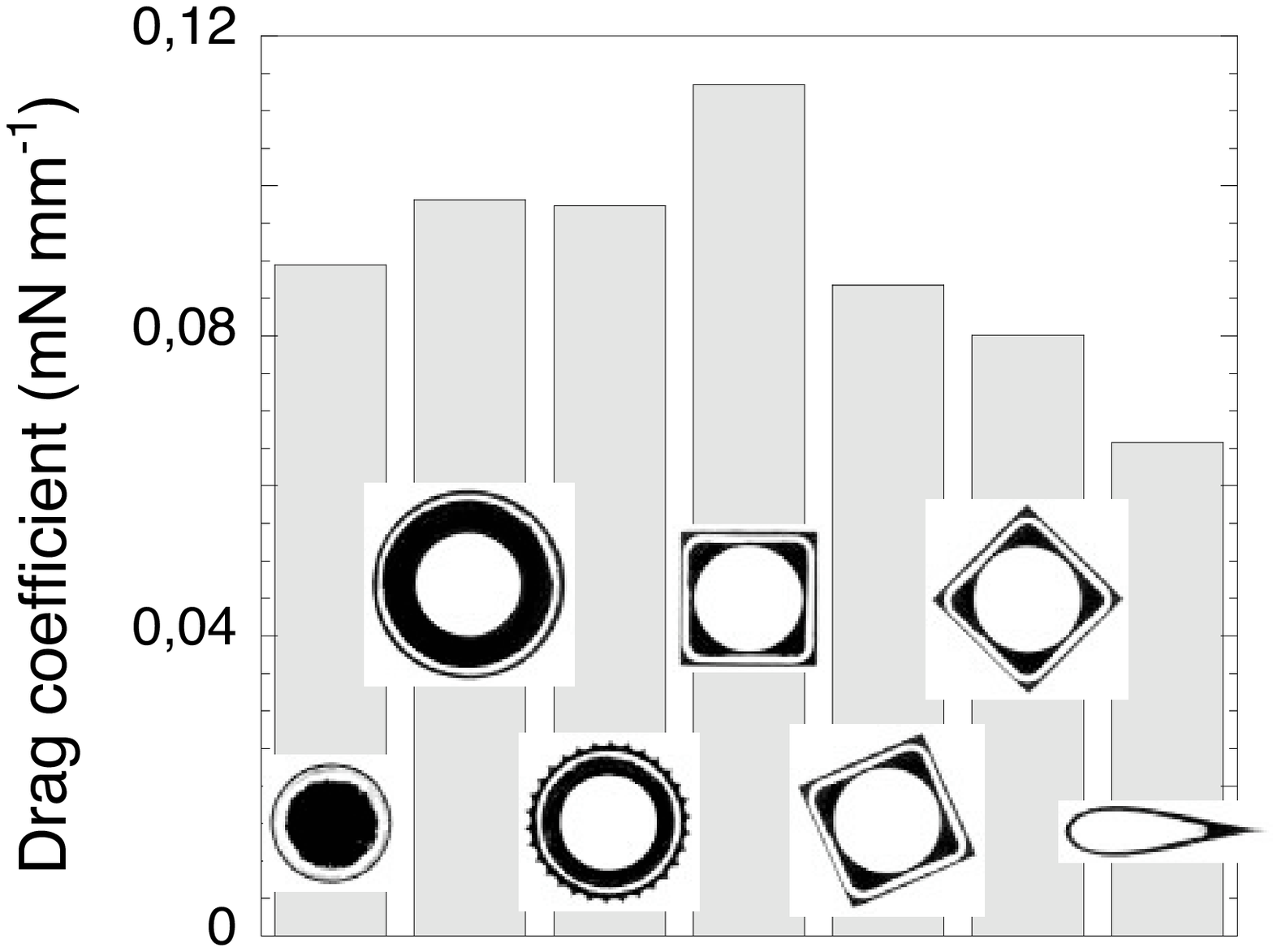}
\caption{\label{Drag_coefficient(obstacle)} Dimensional drag coefficient for
all obstacles. Since the drag exerted on the square does not significantly
depend on its orientation whereas the cross length does, we give the drag
coefficient for the three studied orientations of the square.}\end{center}
\end{figure}

Measurements of drag \emph{versus} flow rate for the five different obstacles
are displayed in Fig. \ref{Drag(Q,obstacle)}. Here again, all data are well
linearly fitted, and as expected, the drag increases with the size of the
obstacle. More quantitative comparison of the obstacles is not straightforward,
since not only their size, but also their shape and boundary conditions, vary.
To investigate the role of all these parameters, we report the viscous
coefficient \emph{versus} the yield drag for the five obstacles, and do a
linear fit passing through zero of all the data (Fig.
\ref{Viscous_coeff(yield_drag)}). This enables to compare the respective
magnitude of elastic and viscous contribution to the drag, and to define an
effective drag $F_{\mathrm{eff}}$ as the orthogonal projection of the data
under linear fit: $F_{\mathrm{eff}} = (m+AF_0)/2A$, where $A = (1.81\pm 0.08)
\times 10^{-3}$ min ml$^{-1}$ is the slope of the linear fitting line. We also
define a dimensional drag coefficient, is units of mN mm$^{-1}$, as the ratio
of the effective drag and the transverse length (orthogonal to the flow), in
analogy with the dimensionless drag coefficient usually defined in
aerodynamics, proportional to the drag and inversely proportional to the cross
section and the velocity of the flow \cite{Comolet1976}. The values of viscous
coefficient, yield drag, their ratio and the dimensional drag coefficient are
displayed in table \ref{ValuesObstacles}, and the values for the dimensional
drag coefficient are displayed as histograms in Fig.
\ref{Drag_coefficient(obstacle)}.

\begin{figure}
\begin{center}
\includegraphics[width=7cm]{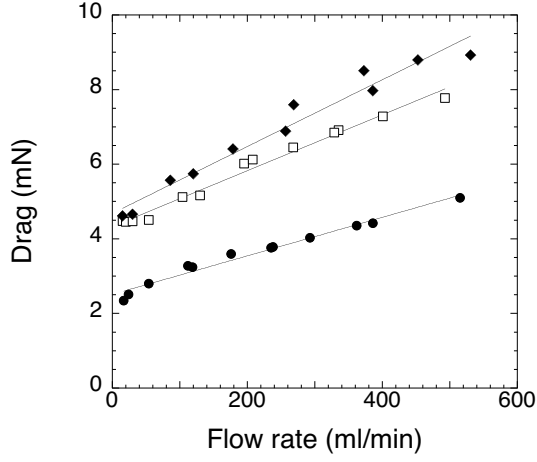}
\caption{\label{Drag(Q,obstacle)} Drag \emph{versus} flow rate, for the
cylinder of diameter 30.0 mm ($\bullet$) and 48.0 mm ($\square$), the cogwheel
($\blacklozenge$), the square ($\vartriangle$) and the airfoil
($\blacktriangledown$). The straight lines are linear fits of the data. The
bulk viscosity is 1.06 mm$^2$ s$^{-1}$ and the bubble area is 16
mm$^2$.}\end{center}
\end{figure}

\begin{table*}
\begin{tabular}{|c|c|c|c|c|c|}
\hline obstacle & cylinder $\varnothing$ 30 mm & cylinder $\varnothing$ 48 mm &
cogwheel & square &
airfoil \\
\hline $F_0$ (mN) & $2.5\pm 0.1$ & $4.6\pm 0.1$ & $4.3\pm 0.1$ & $4.0\pm
0.2$ & $0.5\pm 0.1$ \\
\hline $m$ (mN min l$^{-1}$) & $5.2\pm 0.3$ & $8.6\pm 0.4$ & $7.5\pm 0.3$ &
$6.7\pm
0.5$ & $2.0\pm 0.3$ \\
\hline $m/F_0$ (min l$^{-1}$) & $2.1\pm 0.2$ & $1.9\pm 0.1$ & $1.7\pm 0.1$ &
$1.7\pm
0.2$ & $3.7\pm 1.2$ \\
\hline $C_x$ (mN mm$^{-1}$) & $0.089\pm 0.006$ & $0.098\pm 0.006$ & $0.097\pm 0.005$ & * & $0.066\pm 0.013$ \\
\hline
\end{tabular}
\caption{\label{ValuesObstacles}Yield drag $F_0$, viscous coefficient $m$,
ratio $m/F_0$ and dimensional drag coefficient $C_x$ for each obstacle. The
star symbol recalls that the drag coefficient for the square depends on its
orientation: the value of this coefficient expressed in mN mm$^{-1}$ is
$0.113\pm 0.010$ for an orientation angle of 0$^\circ$, $0.087\pm 0.010$ for an
angle of 22.5$^\circ$ and $0.080\pm 0.005$ for an angle of 45$^\circ$.}
\end{table*}

\section{Discussion}

\subsection{Comparison of our measurements with existing work}

To our knowledge, our work is the first to provide systematic measurements of
the drag exerted by a flowing foam in a channel around an obstacle. This is to
compare to the simulations of Mitsoulis and coworkers
\cite{Zisis2002,Mitsoulis2004}, who computed the drag exerted by a flowing
Bingham plastic past a cylinder similar in geometry to our circle, for
different values of obstacle diameters. A Bingham plastic is characterized by
its yield stress $\tau_y$ and its plastic viscosity $\mu$, and it follows the
constitutive equation: $\tau = \tau_y + \mu\dot{\gamma}$ for $|\tau| > \tau_y$,
and $\dot{\gamma} = 0$ for $|\tau| < \tau_y$, where $\tau$ is the shear stress
and $\dot{\gamma}$ the applied strain. To summarize, Mitsoulis and coworkers
show that the drag exerted by a flowing Bingham plastic around a cylinder
strongly depends on the Bingham number $\mathrm{Bn}=2R\tau_y/\mu V$ comparing
elastic and viscous contribution: at a given Bingham number of order unity,
there is a crossover between a Newtonian-like behaviour of the drag (for
$\mathrm{Bn} \ll 1$) given by formula (\ref{Faxen}), and an elastic-like (for
$\mathrm{Bn} \gg 1$) where drag does not significantly depend on the velocity
and is roughly proportional to the cylinder diameter. Though the validity of
modeling foam as a Bingham plastic is an open debate, this work provides an
interesting comparison to our experimental measurements, for which we now
evaluate the order of magnitude of the Bingham number. The yield stress for a
foam is of order \cite{Princen1983} $0.5\gamma/a$, with $\gamma=31$ mN m$^{-1}$
the surface tension and $a\approx \sqrt{16/(3^{3/2}/2)}\approx 2.5$ mm the
typical length of a bubble edge (we recall that the bubble area is 16.0 mm$^2$
in the considered experiments, and compute $a$ for a hexagonal bubble), so
$\tau_y \approx 6$ Pa (to be rigorous, this overestimates the yield stress for
a wet foam). Furthermore, we can deduce from the value of the viscous
coefficient ($m = 5\times 10^{-6}$ N min ml$^{-1}$ after Fig. \ref{Area}(b)) a
rough value of the plastic viscosity of the foam: dimensional analysis yields
$\mu \approx mS/R$ where $S$ is the cross-section of the foam, so Bingham
number writes $\mathrm{Bn} \approx 2R^2 \tau_y/mQ$. The typical value of flow
rate in our experiments is $10^2$ ml min$^{-1}$, hence the typical Bingham
number equals $\mathrm{Bn} \approx (2\times 0.015^2\times 6)/(5\times
10^{-6}\times 10^2) \approx 5$. Though this is a very rough evaluation, it
tends to show that in our range of flow rates, the Bingham number remains of
order unity, hence both elastic and fluid properties of the foam are involved
in the interaction with the obstacle to create the drag. This corroborates the
measurements of drag in Fig. \ref{Drag(Q,obstacle)} for which elastic and
plastic contribution are of same order of magnitude.

\subsection{Influence of bulk viscosity}

Our measurements of drag \emph{versus} viscosity $\nu$ and flow rate $Q$ yield
the following scaling:
\begin{equation}\label{F(Q,nu)}
  F(Q,\nu) = F_0 + \mathrm{const}\times \nu^{0.77\pm 0.05}Q  ,
\end{equation}
(see section \ref{SectionViscosity}). To our knowledge, this is the first time
that such a scaling is proposed to quantify the dynamical regime of flowing
foams. Up to now, the dynamic regime of flowing foam has been mainly
investigated through the study of pressure drop of foam confined in capillaries
(see Ref. \onlinecite{Cantat2004} and references therein), to model the
behaviour of foams in porous media \cite{Zitha2003,Xu}. Since the seminal work
of Bretherton \cite{Bretherton1961}, who studied the friction between an
infinitely long bubble and a solid wall, all these studies emphasize the role
of the capillary number $\mathrm{Ca}=\eta V/\gamma$, where $\eta$ is the
dynamic bulk viscosity, $\gamma$ its surface tension and $V$ the velocity of
the flowing foam. In the frame of our study, the capillary number is
proportional to the product $\nu Q$. It appears from our scaling
(\ref{F(Q,nu)}) that such a number is not sufficient to describe the dynamic
regime of a flowing foam, because the exponents for viscosity and flow rate
differ significantly, and pressure drop measurements confirm this observation
\cite{Dollet}. Since the velocity-dependent part of the drag is related to
friction of slipping bubbles along the obstacle, Bretherton's theory is
therefore not sufficient to explain our measurements: additional physical
ingredients are involved, like detailed bubble shape and interfacial rheology
(surface elasticity and viscosity). This has not been investigated yet.
Discrepancies from Bretherton's theory have already been widely pointed out and
studied for bubbles and foams in capillaries (see Ref. \onlinecite{Kornev1999}
for a review), but they still considered the capillary number as the essential
dimensionless parameter.

Let us notice that the scaling (\ref{F(Q,nu)}) is a consequence of the chosen
fit (\ref{Linear fit}). We are aware that some rheological studies
\cite{Mason1995,Mason1996,Saint-Jalmes1999} show that storage and loss moduli
of foams happen to depend on the applied shear. This would lead to a behaviour
like $F = F_0 + mQ^\alpha$, the exponent $\alpha$ accounting either for
shear-thinning ($\alpha<1$) or shear-thickening ($\alpha>1$). If such effects
exist in our system, they are small enough to yield results consistent with
$\alpha=1$ within our experimental accuracy. We will thus neglect
shear-thinning or shear-thickening in our further discussion.

\subsection{\label{DiscussionArea}Influence of bubble area}

\subsubsection{Yield drag}

The yield drag has two contributions: an elastic one arising from the elastic
stresses in the network of bubbles, and another one arising from the
anisotropic pressure distribution in the bubbles surrounding the obstacle, as
already shown in preliminary simulations of our experiments \cite{Cox}.

As mentioned in section \ref{ControlParameters}, the depth of the bubbles
adjust to pressure variations. At constant bubble volume, there is therefore a
relation between bubble area and pressure, that we can use to evaluate the
order of magnitude of the pressure contribution to the yield drag. To establish
this relation, we assume that each bubble has the same volume $V_0$, which is
reasonable in our experiments. As a crude model, we treat bubbles as cylinders
of height $h$ and of horizontal area $A$; hence, $V_0 = A_0 h_0 = Ah$, where
$A_0$ and $h_0 = 3.5$ mm refer to mean values. We then assume that the pressure
$P$ inside the bubbles equilibrates with the pressure in the bulk solution in
contact. At vanishing flow rate, this pressure is hydrostatic, hence we write
$P-P_0 = \rho gh = \rho gA_0 h_0/A$, where $P_0$ is a constant reference
pressure, and $\rho = 1.0 \times 10^3$ kg/m$^3$ is the volumetric mass of the
solution. The pressure resultant on the obstacle then writes: $\mathbf{F}_P =
-\iint P\, \mathrm{d}\mathbf{S}$, the integral being taken on the contact
surface between the obstacle and the bubbles. Since $P_0$ is constant,
$\mathbf{F}_P = -\rho gA_0 h_0 \iint \mathrm{d}\mathbf{S}/A$, and
$\mathrm{d}\mathbf{S} = h\, \mathrm{d}\ell \,\mathbf{n}$, where
$\mathrm{d}\ell$ is the length element on the boundary of the obstacle and
$\mathbf{n}$ the normal vector. Since $h = A_0 h_0/A$ one obtains :
\begin{equation}\label{Pressure}
\mathbf{F}_P = -\rho gA_0^2 h_0^2 \oint \frac{\mathrm{d}\ell
\,\mathbf{n}}{A^2}.
\end{equation}
This formula links the pressure contribution to yield drag to the bubble area
field.

\begin{figure}
\begin{center}
\includegraphics[width=8cm]{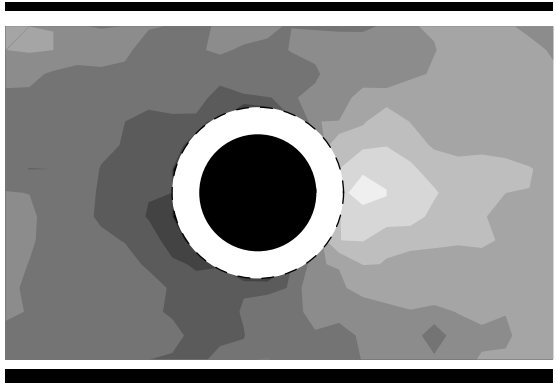}
\caption{\label{AreaField} Bubble area field around the circle of diameter 30
mm. The observed zone is the same as Fig. \ref{Image}. The solution viscosity
is 1.06 mm$^2$ s$^{-1}$, and the mean bubble area is $A_0 = 16.0$ mm$^2$. The
flow rate, 24 ml min$^{-1}$, was chosen such that the velocity-dependent
contribution to the drag is negligible (see Fig. \ref{Drag (area)}). Black
zones represent the obstacle and the channel walls, and white zones the regions
where the bubble area is not measurable precisely. The darker the color, the
lower the bubble area hence the higher the pressure. The area variation is
significant, with a maximum relative variation of 18\%. The pressure is maximal
at the leading side of the circle and minimal at its trailing side (maximal
variation: 70 Pa).}\end{center}
\end{figure}

We illustrate this measurement of pressure on one example (Fig.
\ref{AreaField}). The bubble area field clearly shows the influence of the
obstacle: bubbles are compressed upstream and relax downstream, which
qualitatively shows that the pressure resultant acts in the same sense as
elastic stress. Computing formula (\ref{Pressure}) over the dashed contour in
Fig. \ref{AreaField}, which is the closest contour to the obstacle where bubble
area is properly evaluable, yields an order of magnitude of 0.7 mN for $F_P$,
which is about 30\% of the yield drag (2.5 mN for the studied example, see Fig.
\ref{Area}(a)). The calculation of pressure for various bubble areas, as well
as for higher flow rates and other obstacles, is still in progress, but the
variation depicted in Fig. \ref{AreaField} does not vary qualitatively, and
pressure contribution to yield drag is not negligible.

Another difficulty arises from the variation of fluid fraction with bubble
area. In our setup, the monolayer of bubbles is in contact with a reservoir of
water, and the amount of water in the Plateau borders and films between bubbles
is freely chosen by the system. Therefore, the mean fluid fraction should vary
with bubble area; detailed measurements of this quantity are in progress (first
rough estimate: about 9\%). Furthermore, local effects such as dilatancy
\cite{Weaire2003} could increase the fluid fraction near the obstacle, because
of the shear experienced by the foam in this zone. This complicates the
interpretation of the evolution of yield drag with bubble area, since many
studies have shown that rheological properties of foams and emulsions depend on
fluid fraction \cite{Mason1995,Mason1996,Saint-Jalmes1999}. However, we can
check that the order of magnitude of our measured yield drag agrees
qualitatively with the known value of the yield stress, of order
\cite{Princen1983} $0.5\gamma/a$. Hence, the order of magnitude of the elastic
contribution to the yield drag is $F_{\mathrm{el}} \approx \pi Rh_0 \gamma/a$.
For an hexagonal bubble, $a=\sqrt{2A/3^{3/2}}=0.62\sqrt{A}$, hence
$F_{\mathrm{el}} \approx 5Rh_0 \gamma/\sqrt{A}$, and numerically:
$F_{\mathrm{el}} \approx 8/\sqrt{A}$, with mN and mm$^2$ as units for the force
and for the area. This elastic drag is plotted in Fig. \ref{Area}; though it is
a very rough evaluation, we check that it is of the same order of magnitude as
the yield drag, but that it is not high enough to fit the experimental results:
this is again a signature of the significance of the pressure contribution.

\subsubsection{Viscous coefficient}

We now propose a qualitative argument to explain why the viscous coefficient
decreases with the bubble area, based on the dissipation model of Cantat and
coworkers \cite{Cantat2004}. These authors state that dissipation in flowing
foam is localised in the Plateau borders between bubbles and walls. Hence, the
viscous coefficient should increase with the number of bubbles surrounding the
obstacle, and therefore should decrease with the bubble area, which is actually
seen in Fig. \ref{Area}(b). Note that this model does not capture the increase
observed for the bubble area of 39.3 mm$^2$, but as seen in Fig.
 \ref{Drag (area)}, the drag does not depend affinely on the flow rate for this
area, hence our linear fit is not relevant. As an additional remark, friction
in the foam should strongly depend on the boundary conditions at the interfaces
between films and bubbles, hence the viscous coefficient probably changes with
the surface rheology. It would thus be interesting to investigate the influence
of the surfactant used on the drag measurements.

\subsection{Influence of the obstacle geometry}

\subsubsection{Orientation}

We have shown (Fig. \ref{Drag_on_square(orient)}) that the drag exerted on the
square obstacle does not significantly depend on its orientation. The same
result holds in low Reynolds hydrodynamics, merely owing to the linearity of
the Stokes equation and to the high symmetry of the square \cite{Guyon2001}. On
the other hand, the drag does depend on the orientation at high Reynolds number
\cite{Comolet1976}. We thus think this result provides a good test to validate
possible constitutive equations for foams; it tends to prove the relevance of
linear models, at least in the studied range of control parameters.

\subsubsection{Size, shape and boundary conditions}

We have chosen to compare the various obstacles through an effective drag and a
ratio between the viscous coefficient and the yield drag. We think this is
relevant since this way of comparison involves both the elastic and the viscous
contribution to the drag, which have comparable weight in the studied range of
flow rate (Fig. \ref{Drag(Q,obstacle)}). Furthermore, this provides a way to
compare obstacles of different shapes.

Fig. \ref{Drag_coefficient(obstacle)} shows that the dimensional drag
coefficient does not vary much with the obstacle, except for the airfoil.
Though the cross length is not the unique characteristic length of the
obstacles, this shows that the drag is roughly proportional to the size of the
obstacle. This is not an obvious result: considering the flow of a Newtonian
fluid around a cylinder in the same geometry as ours, and defining like before
a drag coefficient as the ratio between the drag (\ref{Faxen}) and the radius
of the cylinder, it can be shown that this drag coefficient would increase
significantly with the radius. The complete formula (\ref{Faxen}), not shown
for the sake of simplicity (see Ref. \onlinecite{Faxen1946}), yields a drag
coefficient 2.6 times higher for a cylinder of diameter 48 mm than for the one
of diameter 30 mm, whereas the values of Table \ref{ValuesObstacles} show that
the drag coefficients for these two cylinders are comparable in our
experiments. This proves again the significance of elastic effects in our case,
and agrees qualitatively with the results of Mitsoulis \cite{Mitsoulis2004} who
showed that for a Bingham plastic, the effect of channel walls remains weak,
even when the diameter of the cylinder equals the half of the channel width, as
far as elastic effects are dominant.

The ratio between viscous coefficient and yield drag, whose values are
tabulated in Table \ref{ValuesObstacles}, does not change significantly between
the cylinders, the cogwheel and the square, whereas it increases much for the
airfoil. This is clearly a signature of shape: one intuitively expects elastic
effects to act on the cross section orthogonal to the flow to pull the obstacle
streamwise, whereas the viscous contribution to the drag arises from the
friction in the lubrication films between the obstacle and the bubbles slipping
along it. Hence, one expects the viscous contribution to increase with the
cross section parallel to the flow. This explains why the viscous
coefficient/yield drag ratio is higher for the airfoil, owing to the great
difference between the two considered sections for this profile. Furthermore,
the decrease of the drag coefficient for the airfoil, as well as the variation
of this coefficient with the orientation of the square, shows that the shape of
the obstacles influences the results through streamlining: for a given size,
drag is reduced on an obstacle whose shape is well adapted to the flow, like in
aerodynamics.

The values displayed in Table \ref{ValuesObstacles} show that the boundary
conditions do not affect much the drag: the dimensional drag coefficient is
close to those for the two cylinders, whereas the ratio between viscous
coefficient and yield drag is slightly lower. Actually, the cogwheel and the
trapped bubbles form a closed system during the experiment: no rearrangement of
the trapped bubbles occurs after all the cogs have been filled with bubbles. So
this system behaves as an effective obstacle, but with an external boundary
constituted of bubble edges, instead of a solid boundary. This could explain
the slight decrease of the viscous coefficient/yield drag ratio: at low
velocity, the foam feels the presence of the effective obstacle, but at high
velocity, the friction between this effective obstacle and the surrounding
flowing bubbles is lower than the friction between a solid obstacle and its
neighbouring flowing bubbles. To be more quantitative, it would be interesting
to study the influence of interfacial rheology on this friction. Anyway, the
measurements show that the influence of boundary conditions is not dramatic,
probably because it does not change much the features of the flow beyond the
first layer of bubbles.

\section{Conclusions}

This work provides the first detailed and systematic measurements of the force
exerted by a 2D flowing foam on an obstacle as a function of various control
parameters: flow rate, bulk viscosity, bubble volume and obstacle orientation,
size, shape and boundary conditions. All the data show two contributions to the
drag: a yield drag at vanishing flow rate, and a flow rate-dependent
contribution. We have shown that the yield drag is independent of the bulk
viscosity, decreases with bubble volume and linearly increases with the
obstacle size. Moreover, both elastic stresses and pressure contribute
significantly to the yield drag. Fitting the flow rate-dependant contribution
by a linear law, we have shown that the slope (or viscous coefficient)
increases with the bulk viscosity as a power law with an exponent around $3/4$;
moreover, the viscous coefficient globally decreases with the bubble volume and
linearly increases with the obstacle size. Furthermore, we have studied the
influence of the obstacle shape and showed the existence of streamlining
effects in foams, and we pointed out that the effect of boundary conditions on
the obstacle is not striking.

This work opens many perspectives. Other control parameters remain to be
studied, like bubble area polydispersity and rheological properties of the
surfactants. The effects of those parameters on the drag could help to study
their influence on foam rheology. Pressure drop measurements, allowing to study
dissipation in foams \cite{Cantat2004}, are in progress \cite{Dollet}. Now, a
local analysis of the stresses, deformations \cite{Aubouy2003} and velocity
fields is required to provide a more detailed comprehension of the foam
rheology. Such a study is also in progress. The comparison between this local
analysis and the global properties of the foam, such as our drag measurements,
could provide a way to propose and test constitutive equations for the
mechanics of foams.

\begin{acknowledgements}
The authors would like to thank Franck Bernard, Kamal Gam, Julien Deffayet and
Arnaud Huillier for experimental help, the machine shop of Laboratoire de
Spectrom\'etrie Physique and Patrice Ballet for technical support, and Simon
Cox, Wiebke Drenckhan, Isabelle Cantat and Renaud Delannay for enlightening
discussions.
\end{acknowledgements}

\appendix*

\section{Derivation of formula (\ref{Force})}

\begin{figure}
\includegraphics[width=2cm]{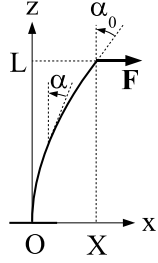}
\caption{\label{Calcul_force} Notations for the calculation of formula
(\ref{Force}).}
\end{figure}

We consider a fiber of vertical length $L$ that experiences a horizontal force
$F$ (Fig. \ref{Calcul_force}). All lengths are adimensionalised by
$\sqrt{IE/2F}$, where $I=\pi D^4/64$ is the inertia moment ($D$ being the fiber
diameter) and $E$ the Young modulus. The position along the fiber is expressed
as a function of the angle $\alpha$. Therefore, the position of the extremity
of the fiber writes in the general case \cite{Landau1986}:
$$ L = 2\sqrt{\sin\alpha_0} \quad , \quad X = \int_0^{\alpha_0}
\frac{\sin\alpha}{\sqrt{\sin\alpha_0-\sin\alpha}} \mathrm{d}\alpha \quad
\Rightarrow \quad X = \int_0^{\arcsin L^2/4}
\frac{\sin\alpha}{\sqrt{L^2/4-\sin\alpha}} \mathrm{d}\alpha.
$$ This yields an implicit expression between the force and the deflection involving elliptic
functions, which is not easy to evaluate.

The fiber can experience large deflections (up to 12 mm for a length of 34.8
mm), so we need a more accurate expression than the linearized one: $X=L^3/6$.
To do that, we develop the previous expression in power series of $L$, that
yields: $X = \dfrac{L^3}{6} + \dfrac{L^7}{280} + \dfrac{L^{11}}{7392} +
\mathcal{O}(L^{15})$. Going back to dimensionalised lengths and inverting the
series yields the formula (\ref{Force}) linking the force and the deflection.

At the maximal deflection, the ratio $X/L$ reaches a value of 0.345. At such a
ratio, the formula (\ref{Force}) gives a precision of 0.3\% over the force,
while the linearized formula $F = 3\pi ED^4 X/64L^3$ yields an error of 9\%.


\begin{thebibliography}{99}

\bibitem{Larson1999} R. G. Larson, \emph{The Structure and Rheology of Complex Fluids}, Oxford University Press, New York
(1999).

\bibitem{Weaire1999} D. Weaire, S. Hutzler, \emph{The physics of foams},
Oxford University Press, Oxford (1999).

\bibitem{Mason1995} T. G. Mason, J. Bibette, D. A. Weitz, \emph{Phys. Rev.
Lett.} \textbf{75}, 2051 (1995).

\bibitem{Mason1996} T. G. Mason, J. Bibette, D. A. Weitz, \emph{J. Coll. Int.
Sci.} \textbf{179}, 439 (1996).

\bibitem{Cohen-Addad1998} S. Cohen-Addad, H. Hoballah, R. H\"ohler, \emph{Phys. Rev. E} \textbf{57}, 6897 (1998).

\bibitem{Saint-Jalmes1999} A. Saint-Jalmes, D. J. Durian, \emph{J.
Rheol.} \textbf{43}, 1411 (1999).

\bibitem{Reinelt2000} D. A. Reinelt, A. Kraynik, \emph{J. Rheol.} \textbf{44}, 453 (2000).

\bibitem{Sollich1997} P. Sollich, F. Lequeux, P. H\'ebraud, M. E. Cates, \emph{Phys. Rev. Lett.} \textbf{78}, 2020 (1997).

\bibitem{Durian1995} D. J. Durian, \emph{Phys. Rev. Lett.} \textbf{75}, 4780 (1995).

\bibitem{Cates2004} M. E. Cates, P. Sollich, \emph{J. Rheol.} \textbf{48}, 193
(2004).

\bibitem{Hohler2004} R. H\"ohler, S. Cohen-Addad, V. Labiausse, \emph{J.
Rheol.} \textbf{48}, 679 (2004).

\bibitem{Monnereau1999} C. Monnereau, M. Vignes-Adler, in \emph{Foams and Emulsions}, Ed. J.F.
Sadoc and N. Rivier, NATO ASI Series E: Applied Sciences, \textbf{354}, Kluwer,
359 (1999).

\bibitem{Prause1995} B. Prause,
J. A. Glazier, S. Gravina, C. Montemagno, \emph{J. Phys.: Cond. Matt.}
\textbf{7}, L511 (1995).

\bibitem{Lambert2004} J. Lambert, I. Cantat, R. Delannay, G. Le Ca\"er, A.
Renault, S. Ruellan, F. Graner, S. Jurine, P. Cloetens, J. A. Glazier,
proceedings of Eufoam 2004, to appear in \emph{Coll. Surf. A}.

\bibitem{Koehler2001} S. A. Koehler, S. Hilgenfeldt, H. A. Stone, \emph{Europhys. Lett.} \textbf{54}, 335
(2001).

\bibitem{Losche1983} M. L\"osche, E. Sackmann, H. M\"ohwald, \emph{Ber.
Bunsenges. Phys. Chem.} \textbf{87}, 2506 (1983).

\bibitem{Courty2003} S. Courty, B. Dollet, F. Elias, P. Heinig, F. Graner, \emph{Europhys. Lett.} \textbf{64}, 709
(2003).

\bibitem{Abdelkader1998} A. {Abd el Kader}, J. C. Earnshaw, \emph{Phys. Rev. Lett.} \textbf{82}, 2610 (1999).

\bibitem{Pratt2003} E. Pratt, M. Dennin, \emph{Phys. Rev. E} \textbf{67}, 051402
(2003).

\bibitem{Smith1952} C. S. Smith, in \emph{Metal Interfaces}, American Society
for Metals, Cleveland, Ohio, 65 (1952).

\bibitem{Debregeas2001} G. Debr\'egeas, H. Tabuteau, J.-M. di Meglio, \emph{Phys. Rev. Lett.} \textbf{87}, 178305
(2001).

\bibitem{Asipauskas2003} M. Asipauskas, M. Aubouy, J. A. Glazier,
F. Graner, Y. Jiang, \emph{Granular Matt.} \textbf{5}, 71 (2003).

\bibitem{Vaz1997} M. F. Vaz, M. A. Fortes, \emph{J. Phys.: Cond. Matt.}
\textbf{9}, 8921 (1997).

\bibitem{Janiaud} \'E. Janiaud, F. Graner, \texttt{cond-mat/0306590}, to appear in \emph{J. Fluid Mech} (2005).

\bibitem{Cox2000} S. J. Cox, M. D. Alonso, S. Hutzler, D. Weaire, in
\emph{Proceedings of the 3rd Euroconference on Foams, Emulsions and their
Applications}, Ed. P. L. J. Zitha, J. Banhart, G. L. M. M. Verbist, Verl. MIT,
Bremen (2000).

\bibitem{Landau1986} L. D. Landau, E. M. Lifshitz, \emph{Theory of
elasticity}, 3rd edition, Reed, Oxford (1986).

\bibitem{Faxen1946} O. H. Fax\'en, \emph{Proc. Roy. Swed. Acad. Eng. Sci.}
\textbf{187}, 1 (1946).

\bibitem{Cox2002} S. J. Cox, D. Weaire, M. F. Vaz, \emph{Eur. Phys. J. E} \textbf{7}, 311 (2002).

\bibitem{Princen1983} H. M. Princen, \emph{J. Coll. Int. Sci.} \textbf{91}, 160 (1983).

\bibitem{Handbook2003} D. R. Lide, \emph{CRC Handbook of Chemistry and
Physics}, 84th Ed., CRC Press, Boca Raton (2003).

\bibitem{Princen1985} H. M. Princen,  \emph{J. Coll. Int. Sci.} \textbf{105}, 150 (1985).

\bibitem{Comolet1976} R. Comolet, \emph{M\'ecanique exp\'erimentale des
fluides, Tome II}, Masson, Paris (1976).

\bibitem{Zisis2002} T. Zisis, E. Mitsoulis, \emph{J. Non Newt. Fluid Mech.}
\textbf{105}, 1 (2002).

\bibitem{Mitsoulis2004} E. Mitsoulis, \emph{Chem. Eng. Sci.} \textbf{59}, 789
(2004).

\bibitem{Cantat2004} I. Cantat, N. Kern, R. Delannay, \emph{Europhys. Lett.}
\textbf{65}, 726 (2004).

\bibitem{Zitha2003} P. L. J. Zitha, \emph{Transp. Porous Media} \textbf{52}, 1
(2003).

\bibitem{Xu} Q. Xu, W. R. Rossen, to appear in \emph{Coll. Surf. A} (2005).

\bibitem{Bretherton1961} F. B. Bretherton, \emph{J. Fluid Mech.} \textbf{10},
166 (1961).

\bibitem{Kornev1999} K. G. Kornev, A. V. Neimark, A. N. Rozhkov, \emph{Adv.
Coll. Int. Sci.} \textbf{82}, 127 (1999).

\bibitem{Cox} S. J. Cox, private communication.

\bibitem{Weaire2003} D. Weaire, S. Hutzler, \emph{Phil. Mag.} \textbf{83}, 2747
(2003).

\bibitem{Guyon2001} \'E. Guyon, J.-P. Hulin, L. Petit, \emph{Hydrodynamique
physique}, EDP Sciences/CNRS \'Editions, Paris (2001).

\bibitem{Dollet} B. Dollet, F. Elias, C. Quilliet, A. Huillier, M. Aubouy, F.
Graner, in \emph{Proceedings of the 5rd European Conference on Foams, Emulsions
and Applications}, to appear in \emph{Coll. Surf. A} (2005).

\bibitem{Aubouy2003} M. Aubouy, Y. Jiang, J. A. Glazier, F. Graner, \emph{Granular
Matt.} \textbf{5}, 67 (2003).

\end{thebibliography}
\end{document}